\documentclass[aps,prb,preprint,onecolumn,superscriptaddress]{revtex4-2}
\usepackage{amssymb,amsmath,amsfonts}
\usepackage{graphicx} 
\usepackage{float} 
\usepackage{natbib} 
\usepackage{color,soul}
\usepackage[colorlinks=true,bookmarks=false,citecolor=blue,urlcolor=blue]{hyperref}
\usepackage{lmodern} 
\usepackage{soul}

\providecommand{\abs}[1]{\vert#1\vert}
\providecommand{\vect}[1]{{\bf {#1}}}

\begin{document}

\title{Probing optical anapoles with fast electron beams}

\author{Carlos Maciel-Escudero}
\thanks{These two authors contributed equally}
\affiliation{Materials Physics Center, CSIC-UPV/EHU, Donostia-San Sebasti\'{a}n, 20018, Spain}
\affiliation{CIC NanoGUNE BRTA and Department of Electricity and Electronics, EHU/UPV, Donostia-San Sebasti\'{a}n, 20018, Spain}

\author{Andrew B. Yankovich}
\thanks{These two authors contributed equally}
\affiliation{Department of Physics, Chalmers University of Technology, 412 96, Gothenburg, Sweden}

\author{Battulga Munkhbat}
\affiliation{Department of Physics, Chalmers University of Technology, 412 96, Gothenburg, Sweden}
\affiliation{Department of Photonics Engineering, Technical University of Denmark, DK--2800, Kgs. Lyngby, Denmark}

\author{Denis G. Baranov}
\affiliation{Department of Physics, Chalmers University of Technology, 412 96, Gothenburg, Sweden}
\affiliation{Center for Photonics and 2D Materials, Moscow Institute of Physics and Technology, Dolgoprudny 141700, Russia}

\author{Rainer Hillenbrand}
\affiliation{CIC NanoGUNE BRTA and Department of Electricity and Electronics, EHU/UPV, Donostia-San Sebasti\'{a}n, 20018, Spain}
\affiliation{IKERBASQUE, Basque Foundation for Science, 48011 Bilbao, Spain}

\author{Eva Olsson}
\email{eva.olsson@chalmers.se}
\affiliation{Department of Physics, Chalmers University of Technology, 412 96, Gothenburg, Sweden}

\author{Javier Aizpurua}
\email{aizpurua@ehu.eus}
\affiliation{Materials Physics Center, CSIC-UPV/EHU, Donostia-San Sebasti\'{a}n, 20018, Spain}
\affiliation{Donostia International Physics Center DIPC, 20018 Donostia-San Sebasti\'{a}n, Spain}

\author{Timur O. Shegai}
\email{timurs@chalmers.se}
\affiliation{Department of Physics, Chalmers University of Technology, 412 96, Gothenburg, Sweden}

\begin{abstract}

Optical anapoles are intriguing charge-current distributions characterized by a strong suppression of electromagnetic radiation. They originate from the destructive interference of the radiation produced by electric and toroidal multipoles. Although anapoles in dielectric structures have been probed and mapped with a combination of near- and far-field optical techniques, their excitation using fast electron beams has not been explored so far. Here, we theoretically and experimentally analyze the excitation of optical anapoles in tungsten disulfide (WS$_2$) nanodisks using  Electron Energy Loss Spectroscopy (EELS) in Scanning Transmission Electron Microscopy (STEM). We observe prominent dips in the electron energy loss spectra and associate them with the excitation of optical anapoles and anapole-exciton hybrids. We are able to map the anapoles excited in the WS$_2$ nanodisks with subnanometer resolution and find that their excitation can be controlled by placing the electron beam at different positions on the nanodisk. Considering current research on the anapole phenomenon, we envision EELS in STEM to become a useful tool for accessing optical anapoles appearing in a variety of dielectric nanoresonators.
\end{abstract}

\maketitle

\section{Introduction}

Anapoles are particular charge-current distributions giving rise to an optical phenomenon characterized by strong suppression of electromagnetic radiation \cite{Dubovik1990}. This phenomenon is typically understood as the interference between the electromagnetic (EM) field produced by the Cartesian electric multipole with the EM field of the toroidal multipole. When this interference is destructive, the system yields a non-radiating current configuration known as the optical anapole state \cite{Afanasiev1995}. 
Excitation of the anapole in a polarizable nanostructure greatly suppresses its scattering cross-section thus providing invisibility to nanoobjects \cite{Kivshar2015}, which offers promising applications in nanophotonics \cite{Kivshar2019,Savinov2019,koshelev2019nonradiating,Sergey2017,Sergey2019}.
Additionally, optical anapoles concentrate EM fields inside the nanoresonators serving to enhance nonlinear harmonics generation\cite{Maier2016}, four-wave mixing \cite{grinblat2017degenerate}, Raman scattering \cite{baranov2018anapole,green2020optical}, and photothermal nonlinearities \cite{zhang2020anapole}. 

Intense experimental and theoretical efforts have been devoted to identify optical anapole states in different dielectric nanostructures. However, the detection of ideal optical anapoles is complicated and usually requires the suppression of other multipole resonances. Typically, this suppression is achieved by engineering the nanoresonator geometry and by structuring the incident illumination \cite{Sergey2019,Savinov2019,Kivshar2019}. Complementary to far-field characterization, understanding how anapoles are excited by localized probes can be of paramount importance to control and realize the full potential of these non-radiating states. An interesting technique of such kind, to probe anapoles, is Electron Energy Loss Spectroscopy (EELS) in Scanning Transmission Electron Microscopy (STEM) due to its ability to access not only electric dipole modes, but also quadrupoles and higher-order modes that do not, or weakly, couple to far-field radiation.

EELS in STEM employs a tightly focused electron beam that allows for mapping the optical properties of a material with tens of meV energy resolution and down to atomic scale spatial resolution while simultaneously relating this information to the samples precise size, shape and structure \cite{DeAbajo2010,DeAbajo2019}. For example, EELS has been used to characterize and map localized plasmons in different nanostructures \cite{Kociak2007,Hohenester2012,Yankovich2017,Camden2018}, to probe electronic excitations such as excitons in Transition Metal Dichalcogenide (TMDC) materials \cite{Tizei2015}, to predict optical toroidal modes in silver heptamer cavities\cite{Talebi2019}, and to map photonic modes of dielectric silicon nanocavities \cite{Duncan2021}. Recently, EELS has also shown to be a useful technique to resolve strong light-matter interactions with unprecedented spatial and spectral resolution\cite{Konecna2018,Yankovich2019,Haran2020,Kociak2020,Lagos2021}. However, a detailed study of optical anapoles excited by fast electrons is still pending.

Here we present a numerical and experimental analysis of optical anapoles and anapole-exciton hybrids in WS$_2$ nanodisks using EELS.
We first identify optical anapole states in model high-index dielectric nanodisks by calculating the electron energy losses experienced by a focused electron beam traveling in aloof trajectory in the vicinity of the nanodisk. Destructive interference between electric and toroidal Cartesian multipoles induced in the nanodisk manifests as a dip in the electron energy loss (EEL) spectra.  Besides being widely used as an excitonic platform, multilayer WS$_2$ possesses a high refractive index \cite{Davoyan2021} which provides a possibility to excite optical anapole states. To experimentally reveal and map anapole excitations in EELS, we thus fabricate WS$_2$ nanodisks and compare experimental EEL spectra with numerical calculations. By  varying the nanodisk dimensions the anapoles supported by the WS$_2$ nanodisk can be tuned to match the resonance of an excitonic transition of WS$_2$ leading to anapole-exciton hybridization. The emergence of this light-matter interaction regime shows the advantage of using TMDCs instead of other conventional high-index materials. Finally, we show the possibility to spatially map optical anapoles and to control their excitation by placing the electron beam at different positions on the nanodisk, demonstrating the potential of EELS to access these special non-radiating charge-current distributions. 

\section{Results}

\subsection{Theoretical prediction of optical anapoles in EEL spectra}

\begin{figure*}[t!]
\includegraphics[width=.7\columnwidth]{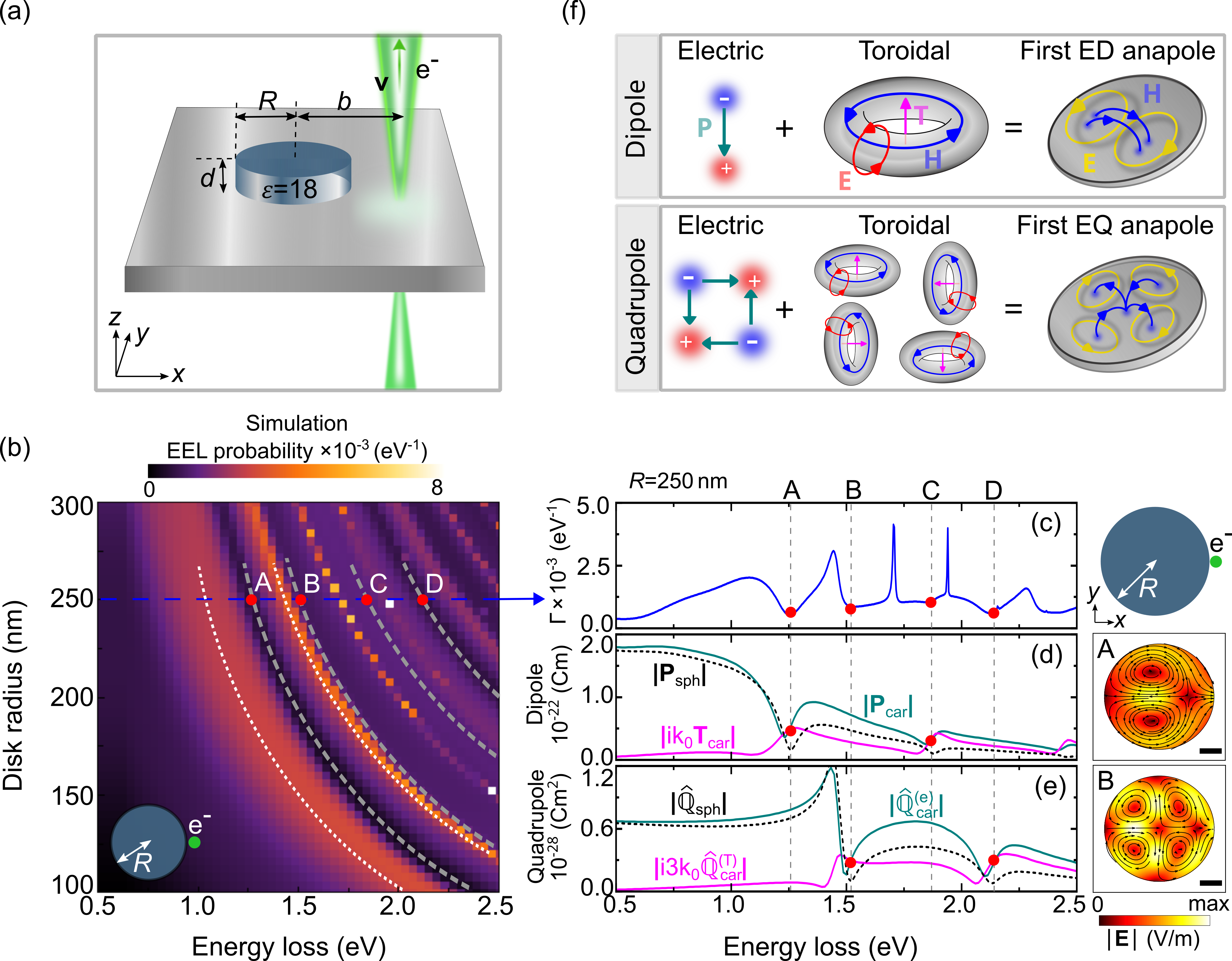}
\caption{\small{Theoretical description of optical anapoles excited by fast electron beams. (a) Sketch of the system under study. The blue cylinder represents the high-index dielectric nanodisk with dielectric function $\varepsilon=18$, thickness $d=55\,\mathrm{nm}$ and radius $R$. The green ray represents the focused electron beam ($e$ is the elementary charge) traveling along the $z$-direction with velocity $v$ at a distance $b$ (impact parameter) with respect to the nanodisk center. The contour plot in (b) shows the EEL probability $\Gamma(\omega)$ calculated as a function of both the nanodisk radius $R$ and the energy loss experienced by the electron beam. White dotted and gray dashed lines are guides to the eye and mark the position of the first two peaks and four dips in the EEL spectra respectively. (c) Simulated EEL probability spectrum (blue line) on a nanodisk with $R=250\,\mathrm{nm}$. The red dots mark the spectral dip positions. (d) Amplitude of the induced dipole moment in the nanodisk. (e) Amplitude of the induced quadrupole moment in the nanodisk. The field plots on the right of panels (d) and (e) depict the amplitude of the total electric field $\abs {\vect{E}(\omega)}$ inside the disk at the plane $z=0$ (half of the nanodisk height), for energies: (A) $1.2575\,\mathrm{eV}$ and (B) $1.52\,\mathrm{eV}$. The scale bar is $100\,\mathrm{nm}$. The maximum value of $\abs{\vect{E}(\omega)}$ in each case is: (A) $5\times 10^8\,\mathrm{V/m}$ and (B) $4\times 10^8\,\mathrm{V/m}$. The inset next to panel (c) illustrates a top view of the nanodisk probed by the electron beam. (f) Sketch of the anapole states formed by the electric and toroidal dipoles and by the electric and toroidal quadrupoles.}}
\label{fig1}
\end{figure*}

We begin our study by describing the general features appearing in the EEL spectra of a model high-index dielectric nanodisk that can exhibit optical anapole states\cite{Shegai2019}. Figure 1(a) illustrates a sketch of the system under study: a dielectric disk is excited by an electron beam traveling in aloof trajectory close to the disk. We choose a disk of variable radius $R$,  thickness $d=55\,\mathrm{nm}$ and permittivity $\varepsilon=18$. The electron beam travels with velocity $v=0.7c$ ($200\,\mathrm{eV}$, $c$ being the speed of light in vacuum) along the $z$-axis at a distance $b=1.1R$ with respect to the nanodisk center (impact parameter). We calculate numerically the EEL probability $\Gamma(\omega)$ using the classical dielectric theory\cite{DeAbajo2010,Hohenester} as specified in Methods and Supplementary Materials S1. 

Figure 1(b) shows the calculated $\Gamma(\omega)$ spectra for different nanodisk radius $R$ in the energy range of $0.5\,\mathrm{eV}$ to $2.5\,\mathrm{eV}$. One can recognize in the spectra peaks (white dotted lines) and dips (gray dashed lines) that monotonously shift to higher energies when the nanoresonator radius $R$ is reduced from $300\,\mathrm{nm}$ to $100\,\mathrm{nm}$.  We associate the peaks to different nanoresonator modes excited by the fast electron beam at the nanodisk (see Supplementary Materials S2). Here we focus on the underlying physics of the spectral dips as they are related to the optical anapole states excited at the disk, as detailed below.

The anapole phenomenon originates from the destructive interference between the electric and toroidal Cartesian moments inside the nanodisk with identical amplitude and far-field patterns. To corroborate anapole excitation with fast electron beams, we thus calculate the EEL probability $\Gamma(\omega)$ for a $250\,\mathrm{nm}$ radius disk (Fig. 1(c)) and perform multipole decomposition of the current density $\vect{J}_{\mathrm{ind}}(\vect{r})$ induced at the nanoresonator by the electron beam. The red points in Fig. 1(c) mark a series of dips revealed in the EEL spectra. The induced current density $\vect{J}_{\mathrm{ind}}(\vect{r})$ can be described by the series of exact multipole moments induced in the disk. In the long-wavelength approximation one can write each multipole moment as a superposition of the so called Cartesian multipole moments \cite{Corbaton2015,Corbaton2018,Corbaton2019}. For example, the spherical electric dipole moment $\vect{P}_{\mathrm{sph}}(\omega)$ in the long-wavelength approximation is described by the superposition of the electric and toroidal Cartesian dipole moments: $\vect{P}_{\mathrm{sph}}(\omega)\approx \vect{P}_{\mathrm{car}}(\omega)+ik_0 \vect{T}_{\mathrm{car}}(\omega)$, where $k_0=\omega/c$ is the wave number in vacuum. When the contributions of the electric and toroidal Cartesian dipole moments are out-of-phase ($\vect{P}_{\mathrm{car}}(\omega)=-ik_0\vect{T}_{\mathrm{car}}(\omega)$), that is, they interfere destructively, the electric spherical dipole moment $\vect{P}_{\mathrm{sph}}(\omega)$ approaches zero. This is the condition for the excitation of optical anapole states \cite{Kivshar2015}. In Supplementary Materials S3, we provide further details on the calculations of the spherical and Cartesian multipoles as well as on the multipole expansion in the long-wavelength approximation. In order to corroborate anapoles excitation in the high-index dielectric disk, we calculate the dipole and quadrupole moments of the induced current density $\vect{J}_{\mathrm{ind}}(\vect{r})$ inside the $250\,\mathrm{nm}$ disk.

Figure 1(d) shows calculated amplitudes of the electric spherical $\abs{\vect{P}_{\mathrm{sph}}(\omega)}$ (black dashed line) dipole moment, as well as the Cartesian electric $\abs{\vect{P}_{\mathrm{car}}(\omega)}$ (green line) and scaled toroidal $\abs{i k_0 \vect{T}_{\mathrm{car}}(\omega)}$ (magenta) dipole moments. From the $\abs{\vect{P}_{\mathrm{sph}}(\omega)}$ spectrum, we recognize a dip at around $1.25\,\mathrm{eV}$ (marked with A). The appearance of the dip is a consequence of the Cartesian dipole and toroidal dipole moments having the same amplitude but opposite phase (green and magenta lines intersect at an energy marked with A in Fig. 1(d)). Most importantly, by comparing the EEL (Fig. 1(c)) and the spherical dipole moment (blacked dashed line in Fig. 1(d)) spectra, we find that the dip in $\abs{\vect{P}_{\mathrm{sph}}(\omega)}$ corresponds to the first dip in $\Gamma(\omega)$, indicated by the fist gray dashed lines in Figs. 1(c) and (d). This corroborates that the first resonance dip revealed in the EEL spectra of the nanodisk belongs to an optical anapole state excited by the fast electron beam. The distribution of the electric field plot at energy $1.2575\,\mathrm{eV}$ around the disk (right panel of Fig. 1(d)) shows the anapole near-field pattern inside the disk, revealing regions of intense fields inside the nanodisk and two opposite vortices typically found in the optical phenomena involving anapoles. We refer to this anapole as the first electric dipole (ED) anapole state. We note that the ideal anapole (zero in the EEL probability) cannot arise due to higher-order multipolar contributions to the energy loss and thus, an ``attenuated dip'' can be observed in the EEL spectra. 

Analogously, we can understand the dip at $1.52\,\mathrm{eV}$ in the EEL spectrum (marked with B in Fig. 1(c)) as due to the destructive interference between electric and toroidal quadrupole current distributions with identical far-field patterns. To demonstrate this, we calculate the electric spherical $\hat{\mathbb{Q}}_{\mathrm{sph}}(\omega)$, Cartesian electric $\hat{\mathbb{Q}}^{(\mathrm{e})}_{\mathrm{car}}(\omega)$ and toroidal  $\hat{\mathbb{Q}}^{(\mathrm{T})}_{\mathrm{car}}(\omega)$ quadrupole moments of the induced current density $\vect{J}_{\mathrm{ind}}(\vect{r})$. In the long-wavelength approximation, one finds that the spherical electric quadrupole moment has the following form $\hat{\mathbb{{Q}}}_{\mathrm{sph}}(\omega) \approx \hat{\mathbb{Q}}^{(\mathrm{e})}_{\mathrm{car}}(\omega)+3ik_0 \hat{\mathbb{Q}}^{(\mathrm{(T})}_{\mathrm{car}}(\omega)$ (see Supplementary Materials S3) and thus, the condition for the excitation of electric quadrupole (EQ) anapole states is $\hat{\mathbb{Q}}^{(\mathrm{e})}_{\mathrm{car}}=-3ik_0 \hat{\mathbb{Q}}^{(\mathrm{T})}_{\mathrm{car}}$. In Fig. 1(e) we show the amplitude of $\abs{\hat{\mathbb{Q}}_{\mathrm{sph}}(\omega)}$ (black dashed line), $\abs{ \hat{\mathbb{Q}}^{(\mathrm{T})}_{\mathrm{car}}(\omega)}$ (green line) and  $\abs{3 i k_0 \hat{\mathbb{Q}}^{(\mathrm{T})}_{\mathrm{car}}(\omega)}$ (magenta line). From the green and magenta spectra we clearly see that the Cartesian and toroidal quadrupole moments sustain the same amplitude but opposite phase at $1.52\,\mathrm{eV}$, which confirms that the dip at $1.52\,\mathrm{eV}$ in the EEL spectra corresponds to the excitation of the first EQ anapole state. We note that the EQ anapole states in the disk can not be excited using plane wave illumination at normal incidence, as we show in Supplementary Materials S4. The difference in the properties of the first ED and the first EQ anapoles excited by the fast electron beam can be observed in the right panel of Fig. 1(e), where we show the near-field pattern induced at $1.52\,\mathrm{eV}$ (marked with B) inside the disk. Notice that the field pattern of the first EQ anapole exhibits two additional vortices compared to the field pattern of the first ED anapole (see right panel of Fig. 1(d)). This increase in the number of vortices can be attributed to the quadrupolar distribution of the induced current density $\vect{J}_{\mathrm{ind}}(\vect{r})$ inside the disk, as sketched in Fig. 1(f). Finally, we note that subsequent dips in the EEL spectrum (marked C and D in Fig. 1(c)) belongs to the excitation of higher-order anapole states such as the second ED anapole (marked C) and the second EQ anapole (marked D) states.

The possibility to probe optical anapole states with fast electron beams turns EELS into a promising tool for fundamental studies of optical phenomena involving anapoles. To experimentally verify our numerical calculations we fabricate high-index TMDC WS$_2$ nanodisks with various radii and performed electron energy loss spectroscopy to spatially resolve their optical behavior.

\subsection{Fabrication of WS$_2$ nanodisks}

We synthesize WS$_2$ nanodisks by transferring a mechanically exfoliated WS$_2$ flake onto a $50\,\mathrm{nm}$ thick SiN membrane and performing a combination of e-beam lithography and dry etching. This process enables the creation of donut-shaped etched patterns with isolated nanodisks at their center with selected radii (see Methods and Supplementary Materials S5 for additional details). The thickness of the nanodisk determines its optical response to a probing fast electron, therefore we experimentally measure the disk thickness using three independent methods: EELS, tilted view STEM imaging, and the combination of optical reflectivity and transfer-matrix fitting (Supplementary Materials S6). From the three methods, we find a disk thickness of around $d\approx70\,\mathrm{nm}$.

\begin{figure}[t!]
\includegraphics[width=\textwidth]{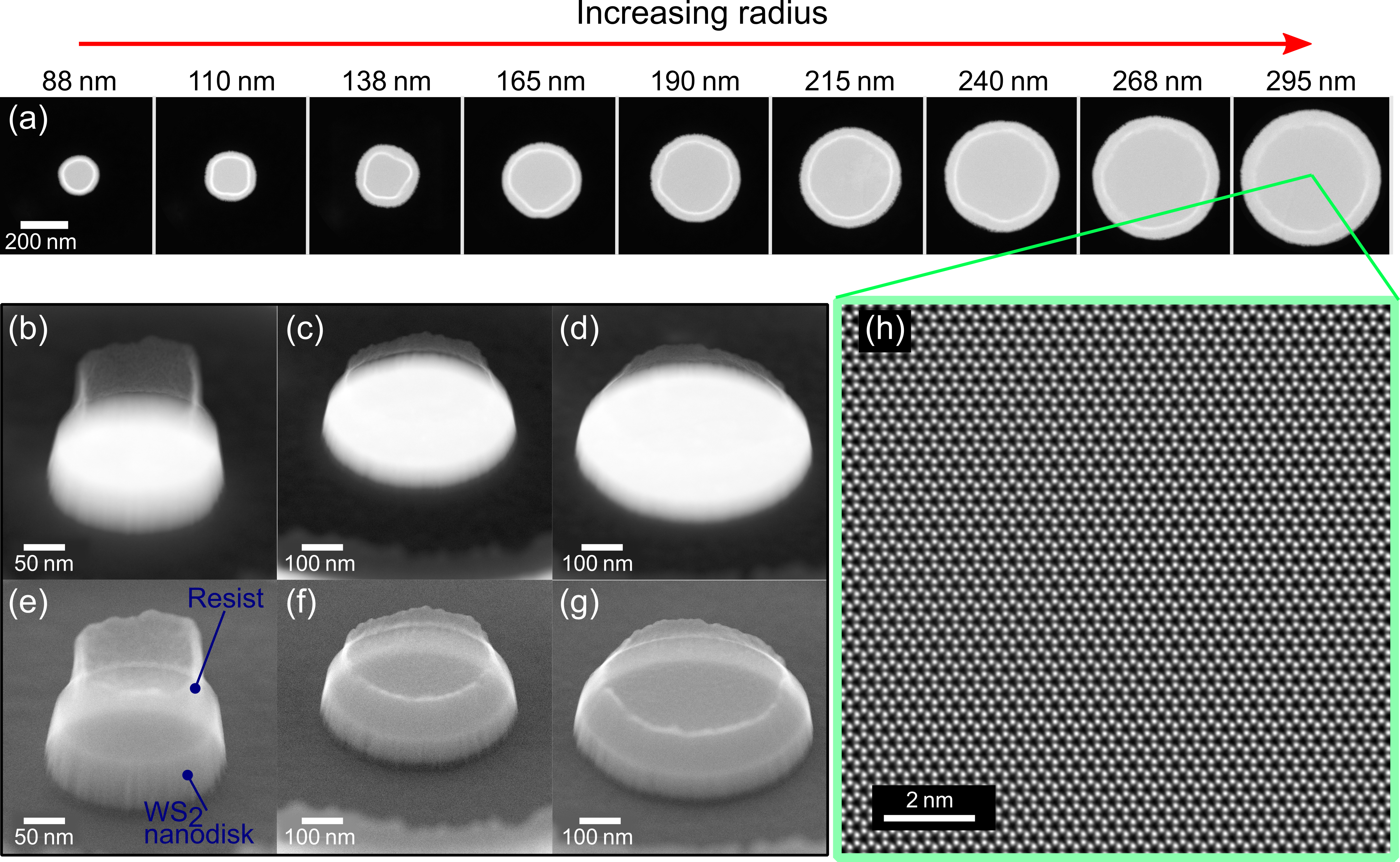}
\caption{\small{Fabrication of WS$_2$ nanodisks. (a) Plan view HAADF STEM images of nine WS$_2$ nanodisks with radii ranging from $88\,\mathrm{nm}$ to $295\,\mathrm{nm}$. Disk radii are shown at the top of each image and a scale bar of $200\,\mathrm{nm}$ stands for all images.  (b)-(d) 53$^{\circ}$ tilted HAADF STEM images of the nanodisks with radii $110\,\mathrm{nm}$, $240\,\mathrm{nm}$ and $295\,\mathrm{nm}$, respectively. (e)-(g) 53$^{\circ}$ tilted SE STEM images simultaneously acquired of the three nanodisks shown in panels (b)-(d). (h) Atomic resolution HAADF STEM image taken from the center of the largest ($295\,\mathrm{nm}$) disk.}}
\label{fig2}
\end{figure}

In Fig. 2(a) we  show high-angle annular dark-field (HAADF) STEM images of nine WS$_2$ nanodisks of different size, demonstrating the ability to precisely synthesize isolated nanodisks with controllable radii ranging from around $88\,\mathrm{nm}$ to $295\,\mathrm{nm}$. Simultaneously acquired $53^{\circ}$ tilted HAADF and secondary electron (SE) STEM images of nanodisks with various radii (Figs. 2(b)-(g)) reveal the morphology of the nanoresonators and the residual resist material that remains on top of the WS$_2$ nanodisks after fabrication, as indicated in Fig. 2(e). High-angle annular dark-field STEM images are dominated by mass-thickness Z-contrast \cite{Pennycook} and thus, the brightest regions reveal the position and shape of the WS$_2$ nanodisks below the primarily low-Z residual resist material. To identify the size and morphology of both the WS$_2$ nanodisks and the residual material, we use SE image contrast which is sensitive to surface topography \cite{Wall2009}. From Figs. 2(b)-(g) we observe two main features of the WS$_2$ nanodisks:  (i) the edges show small vertically aligned surface variations and (ii) a tapered side surface with larger radii at the base. These variations of the WS$_2$ nanodisk radius are small compared to the average nanodisk radius, and thus modeling our nanodisks as perfect disks is an adequate description of the system. The residual resist could lead to minor shifts in energy of the disk modes, but does not alter the excitation of the nanoresonator modes and anapole states. Therefore, the resist has not been included in our model calculations. We address the reader to Supplementary Materials S7, where we provide further details on the chemical composition of the fabricated samples. In Fig. 2(h) we show an atomic-resolution HAADF STEM image from the center of a nanodisk, revealing that the single crystalline atomic structure of the WS$_2$ flake is preserved after the nanodisk fabrication process.

\subsection{Probing optical anapoles in WS$_2$ nanodisks using EELS}

To experimentally investigate the optical response of WS$_2$ nanodisks to a fast electron beam, we performed monochromated STEM EELS experiments using a $200\,\mathrm{keV}$ electron beam with less than $1\,\mathrm{nm}$ spatial resolution and $20\,\mathrm{meV}-40\,\mathrm{meV}$ energy resolution (see Methods). In Fig. 3(a) we show the collected EEL spectra as a function of nanodisk radius as obtained when an aloof electron beam passes outside ($b<R+5\,\mathrm{nm}$) the edge of each nanodisk shown in Fig. 2(a). From the EEL spectra we observe a low-loss signal composed of multiple sharp peaks and dips (see Fig. 3(a)). The number of peaks and dips decreases steadily as the disk size is reduced and their position shifts monotonously to higher energies in agreement with our theoretical prediction (Fig. 1(b)). Importantly, EEL spectra of the WS$_2$ disks exhibit a spectral peak at around $ 1.95\,\mathrm{eV}$ (bright stripe in Fig. 3(a)) present in all the nanoresonators. Notice that above this energy the EEL signal dampens and blurs. For a better quantitative comparison, we show in Figs. 3(b) and 3(c) the EEL spectra extracted from Fig. 3(a) for the $268\,\mathrm{nm}$ (blue spectrum) and $110\,\mathrm{nm}$ (red spectrum) radius nanodisks, respectively. We clearly distinguish the peak at $1.96\,\mathrm{eV}$ which we attribute to the excitation of the A-exciton absorption band of WS$_2$ (for further details and discussion see Supplementary Materials S1), as identified in the EEL spectrum of an unpatterned WS$_2$ flake shown by the green lines in Figs. 3(b) and (c).

To better understand the measured EEL spectra, we numerically calculate the EEL probability $\Gamma(\omega)$ for different WS$_2$ nanodisks with radius ranging from $88\,\mathrm{nm}$ to $295\,\mathrm{nm}$ (Fig. 3(d)). In contrast to the numerical simulations shown in Fig. 1, we performed these calculations shown in Figs. 3(d)-(f) with a permittivity function of the disk that includes the A-exciton absorption band of WS$_2$ (see bright stripe in Fig. 3(d) and Supplementary Materials S1). We find a good agreement between the experimental and simulated spectra (Figs. 3(a) and 3(d)) in the number, position, and dispersion of peaks and dips across the complete disk radii range (notice that the appearance of the A-exciton in the EEL spectrum for every disk radii is also consistent between experiments and simulations). Above $1.96\,\mathrm{eV}$ the EEL signal is better resolved in the simulated than in the experimental spectra (region between $2\,\mathrm{eV}-2.5\,\mathrm{eV}$ in Figs. 3(a) and 3(d)). We attribute this blurring in the experimental signal to the excitation of B- and C-exciton absorption bands of WS$_2$ that are not included in the simulations.
 
To check whether the dips in the experimental spectra are due to the excitation of optical anapoles, we analyze the EEL signal of the $268\,\mathrm{nm}$ disk radius (blue line in Fig. 3(b)). To that end, we extract from Fig. 3(d) the corresponding simulated spectrum and plot it in Fig. 3(e) (solid blue line). One can observe that both experimental and simulated results show two attenuated dips between $1\,\mathrm{eV}$ and $1.5\,\mathrm{eV}$, marked with A and B in Fig. 3(b) and (e). As we discussed in Fig. 1, these two attenuated dips can be associated to the disk first ED and the first EQ anapole states. We have verified that the dips in this energy range are not caused by the intrinsic losses in WS$_2$. Specifically, we simulate the spectrum of the model disk with $R=268\,\mathrm{nm}$, $d=70\,\mathrm{nm}$ and artificial permittivity $\varepsilon=18$ (mimicking the WS$_2$ without the A-exciton resonance at $1.96\,\mathrm{eV}$). The spectrum is shown in Fig. 3(e) (thin blue line). By comparing the solid and the thin blue lines we can observe that for low energies the first two spectral dips (marked with A and B) appear nearly at the same energy for both types of disk (compare solid and thin blue lines between $0.5\,\mathrm{eV}-1.5\,\mathrm{eV}$). This confirms the excitation of optical anapoles in the WS$_2$ disk by the fast electron beam. We note that the differences between the solid and the thin blue spectra in Fig. 3(e) are a direct consequence of the appearance of the A-exciton resonance at $1.96\,\mathrm{eV}$ (identified in the EEL spectrum of a WS$_2$ flake shown by the green lines in Fig. 3(e)).

\begin{figure}[t!]
\includegraphics[width=0.75\columnwidth]{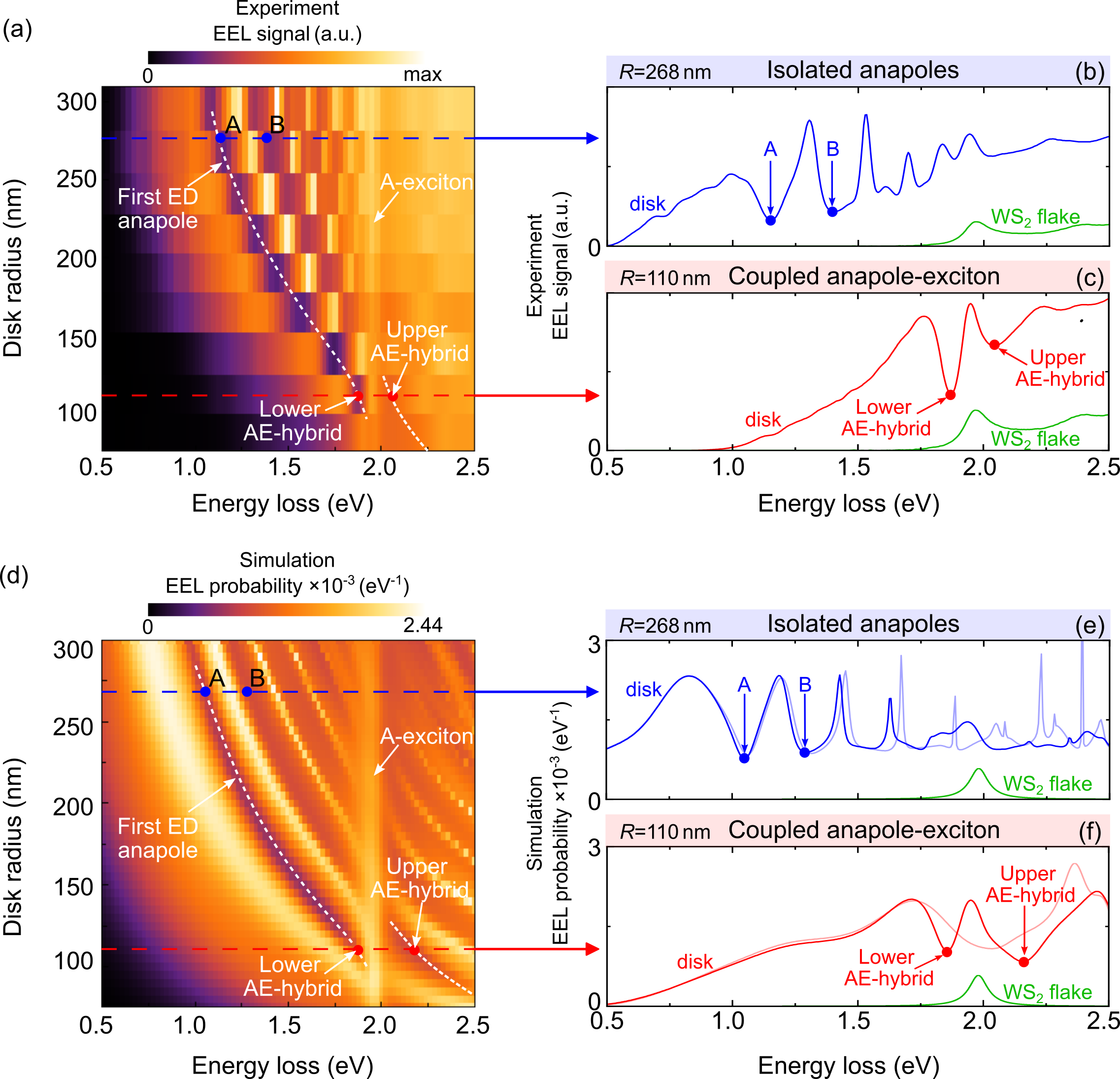}
\caption{\small{Electron energy loss spectra of the WS$_2$ nanodisks. The contour plot in (a) shows the experimental EEL spectra of the nine disks displayed in Fig. 2(a). Blue solid line in (b) and red solid line in (c) show the EEL spectra for disks with $R=268\,\mathrm{nm}$ and $R=110\,\mathrm{nm}$, respectively. Green lines in panels (b) and (c) show EEL spectra from an unpatterned WS$_2$ flake. The contour plot in (d) shows the simulated EEL probability $\Gamma(\omega)$ as a function of both the nanodisk radius $R$ and the energy loss experienced by the electron beam. Blue solid line in (e) and red solid line in (f) show the EEL spectra for disks with $R=268\,\mathrm{nm}$ and $R=110\,\mathrm{nm}$, respectively. Thin blue and thin red spectra correspond to $\Gamma(\omega)$ obtained with disks of artificial permittivity $\varepsilon=18$. Green lines in panels (e) and (f) show calculated $\Gamma(\omega)$ for a WS$_2$ flake of $70\,\mathrm{nm}$ thickness (see Supplementary Materials S1).  White dashed lines in panels (a) and (d) are guides to the eye indicating anti-crossing of the first ED anapole and the A-exciton. Green curves in (a) and (b) were scaled to be consistent with the relatives heights in (e) and (f).}}
\label{fig3}
\end{figure}

\subsubsection{Anapole-exciton hybridization}

The coexistence of an excitonic resonance and the dispersive anapoles in the same object allows these resonant features to couple and hybridize with each other when varying the nanodisk radius $R$. To explore this aspect, we trace the first ED anapole state upon decreasing the nanodisk radius from $268\,\mathrm{nm}$ to $110\,\mathrm{nm}$ (white dashed line in Fig. 3(a) and (d)) until it reaches the energy of the A-exciton resonance, where a splitting (anti-crossing) of the dip is produced. This behavior is clearly shown in Fig. 3(f), where we plot the simulated spectrum of the $110\,\mathrm{nm}$ disk radius (solid red line, extracted from Fig. 3(d)). For comparison, we also plot the spectrum of the model disk with artificial permittivity $\varepsilon=18$ (thin red line). By comparing the spectra of both type of disks, one can observe that the attenuated anapole dip at $2\,\mathrm{eV}$ (thin red line) splits into two dips at $1.86\,\mathrm{eV}$ and $2.16\,\mathrm{eV}$ in the solid red line. Due to the hybrid nature of these dips, we refer to them as the lower anapole-exciton-hybrid (lower AE-hybrid) and the upper anapole-exciton-hybrid (upper AE-hybrid). We can also see a peak in between the two dips which originates by the excitation of excitons that do not couple to the anapole. The splitting of the dips, together with the anti-crossing feature, are signatures of the coupling between the first ED anapole state and the A-exciton, consistent with previous observations in far-field optical spectroscopy of WS$_2$ nanodisks \cite{Shegai2019}.

The anti-crossing observed, and fully identified in Fig. 3(d) resembles the typical situation of coupling between an EM mode confined in an optical cavity and a dipolar excitation. The anapole, however, is not an EM mode of the disk, but instead is the result of interference between at least two resonant modes excited by the electron beam \cite{Joannopopulos2014,monticone2019can}. To explain the coupling between the first ED anapole and the A-exciton, we implement an analytical model of the response of the coupled system based on temporal coupled mode theory (TCMT). Within this framework, we model the anapole-exciton system using an effective $3 \times 3$ Hamiltonian containing the eigenfrequencies of two EM modes, whose far-field interference produces the anapole dip, coupled to a third non-radiating mode representing the A-exciton of WS$_2$. We use this model and apply the following procedure to reproduce the simulated and experimental EEL spectra. First we reproduce the simulated spectra of the model disk with $\varepsilon=18$ and find the eigenfrequencies of the two EM modes producing the first ED anapole dip. We then use these values to reproduce the experimental and simulated EEL spectra of the WS$_2$ disks (Figs. 3(a) and (d)) and find the coupling strengths between each EM mode and the A-exciton. By diagonalizing the effective $3 \times 3$ Hamiltonian of the system, we find the eigenfrequencies of the new hybrid modes. The results obtained with this procedure are shown in Supplementary Materials S8, where we describe in detail the coupled-mode model. The analysis reveals a clear anti-crossing between the hybrid modes as a function of the inverse of the nanodisk radius, $1/R$, indicating that the two electromagnetic modes are strongly coupled to the A-exciton resonance. The two dips that result from the coupling between the two modes and the A-exciton correspond to the lower and upper AE-hybrids.  

\subsection{Real-space mapping of optical anapole states}

An advantage of EELS in STEM is its ability to acquire spectral information of a sample with subnanometer spatial resolution. Typically, this is achieved by scanning the sample area with the fast electron beam, thus obtaining spectral information of the sample at different beam positions (see Methods). In Fig. 4 we apply this capability to spatially resolve disk modes and anapoles states excited in the WS$_2$ disk. Specifically, we collect the EEL signal as a function of the impact parameter $b$. The cylindrical symmetry of the nanodisk along the $z$-axis, together with the trajectory of the probing electron beam, yields an EEL signal that depends only on the impact parameter. Therefore, we represent the spatial distribution of the EEL signals obtained from the WS$_2$ nanodisks by a 2D-EEL line profile showing the energy loss as a function of impact parameters for each energy. 

In Fig. 4 we show experimental and simulated 2D-EEL line profiles for nanodisks with $R=268\,\mathrm{nm}$ (Figs. 4(a) and 4(b)) and $R=110\,\mathrm{nm}$ (Figs. 4(f) and 4(g)). The line profiles for the $R=268\,\mathrm{nm}$ nanodisk reveal a low-loss EEL signal confined to an annular region with a maximum at the edge of the nanodisk (see region between $0.5\,\mathrm{eV}$ to $1.5\,\mathrm{eV}$ in Figs. 4(a) and Fig. 4(b)). We show in Fig. 4(e) the simulated spectrum for a beam that is close to the nanodisk edge (blue dashed line in Fig. 4(a)). We recognize the fist ED and EQ anapole dips at around $1\,\mathrm{eV}$ and $1.25\,\mathrm{eV}$, respectively. The electric field distribution inside the nanodisk associated to these energies (see Figs. 4(c) and 4(d)) corroborate the nature of these dips. Interestingly, when the electron beam passes through the nanodisk center ($b=0$) the EEL signal becomes nearly zero, as shown in Figs. 4(a) and Fig. 4(b). In this case, the electron beam is not able to efficiently excite the disk modes due to the cylindrical symmetry of the disk.

\begin{figure}[t!]
\includegraphics[width=0.75\columnwidth]{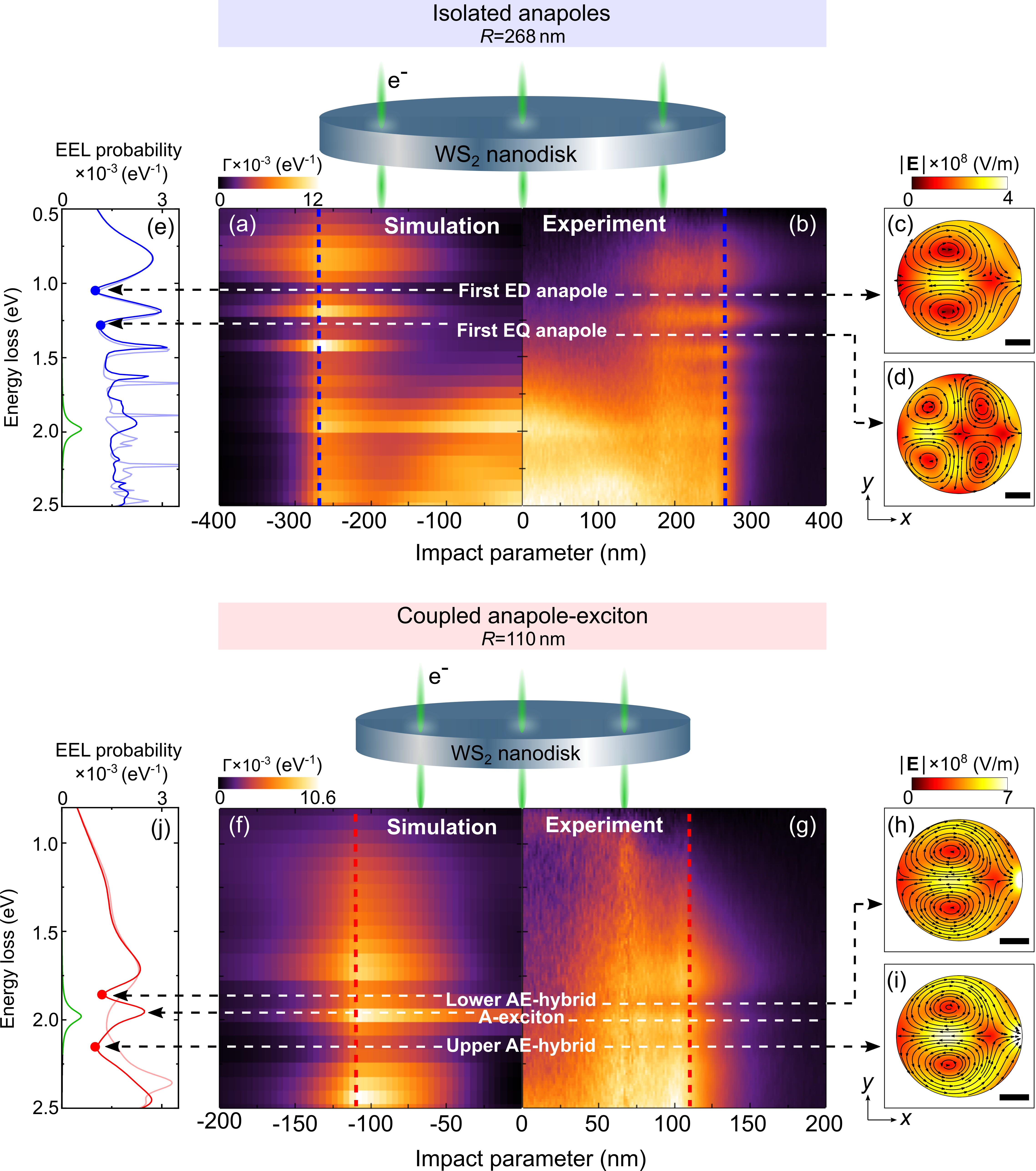}
\caption{\small{Spatially-resolved EEL of WS$_2$ nanodisks. (a) Simulated and (b) experimental EEL spectra recorded along the impact parameter $b$ of a disk with $268\,\mathrm{nm}$ radius as depicted in the schematics above the spectra. Blue dashed lines indicate $b=268\,\mathrm{nm}$. White dashed lines indicate the first ED (at $1.048\,\mathrm{eV}$) and the first EQ (at $1.276\,\mathrm{eV}$) anapole states. Panels at the right of (b) show the amplitude of the total electric field $\abs {\vect{E}(\omega)}$ at the plane $z=0$ for the energies (c) $1.048\,\mathrm{eV}$ and (d) $1.276\,\mathrm{eV}$. The scale bar is $100\,\mathrm{nm}$. (e) Simulated $\Gamma(\omega)$ spectra for the WS$_2$ disk (solid blue line), the model disk with $\varepsilon=18$ (thin blue line) and a WS$_2$ flake of $70\,\mathrm{nm}$ thickness. (f)-(j) same as (a)-(e) but for the $R=110\,\mathrm{nm}$ nanodisk. The scale bars in panels (h) and (j) are $50\,\mathrm{nm}$. We used for the calculation of $\Gamma(\omega)$ an impact parameter equal to $b=1.05R$.}}
\label{fig4}
\end{figure}

Finally, the line profile obtained from the $R=110\,\mathrm{nm}$ nanodisk reveals an EEL signal that is spatially confined to an annular region from half the disk radius to significantly outside of the nanodisk. The calculated electric field distributions at the spectral dips (around $1.75\,\mathrm{eV}$ and $2.25\,\mathrm{eV}$ as indicated in Fig. 4(j)) display a clear ED anapole-like field pattern (Figs. 4(h) and 4(i)). These field distributions reveal that the first ED anapole is hybridized with the A-exciton absorption band (bright stripe around $1.95\,\mathrm{eV}$) to produce the anapole-exciton hybrids. These results open up the possibility for future EELS experiments on systems exhibiting more complicated spatial behavior of the isolated anapoles or anapole-exciton hybrids by, for example, breaking the cylindrical symmetry of the system.

\section{Conclusion}

In summary, we demonstrated that electron energy loss spectroscopy can be applied to probe optical anapole states in high-index dielectric nanoresonators. To that end, we calculated the electron energy loss probability of high-index dielectric nanodisks and showed that the prominent dips in the the EEL spectra are associated with the excitation of optical anapoles in the disk. We experimentally verified our theoretical prediction by performing EELS of WS$_2$ nanodisks and revealed optical anapoles and anapole-exciton hybrids excitation within the same nano object.  Additionally, we demonstrated that EELS in STEM allows for spatial mapping of WS$_2$ nanodisk modes, isolated anapoles and  anapole-exciton hybrids with subnanometer resolution. By locating the electron beam at specific positions along the WS$_2$ nanodisk, we can effectively control the modes excitation and thus the formation of the optical anapole. 

Our results show that EELS in STEM is a powerful tool to study dark scattering states and their complex interactions with the electronic structure of dielectric materials beyond what is possible using optical techniques. We anticipate our results will enable new possibilities for studying of higher-order and magnetic anapole states in dielectric nanoresonators with subnanometer spatial resolution.

\section*{Methods}

\subsection{Sample fabrication}

The multilayer WS$_2$ ($\sim 70\,\mathrm{nm}$) flake was mechanically exfoliated from bulk crystal (HQ graphene) onto polydimethylsiloxane (PDMS) stamps using the scotch-tape method, and then transferred onto a $50\,\mathrm{nm}$ thick silicon nitride (SiN) membrane TEM grid (TEMwindows.com) with an all dry-transfer method. The fabrication of WS$_2$ nanodisks was achieved using our previously developed method with a combination of e-beam exposure of a positive resist and dry etching \cite{munkhbat2020transition,munkhbat2022nanostructured,Shegai2019}. In brief, nanopatterning WS$_2$ ($\sim 70\,\mathrm{nm}$) flake into the nanodisks was carried out by first spin coating a positive ARP 6200.13 resist at $4000\,\mathrm{rpm}$ for $1\,\mathrm{min}$, followed by soft-baking at $120^{\circ}\,\mathrm{C}$ for $5\,\mathrm{min}$. To fabricate the nanodisks on a thin SiN membrane by dry etching, the sample was exposed with donut-shaped patterns with various inner and outer diameters using a Raith EPG5200 electron beam lithography system ($100\,\mathrm{keV}$) to make a resist mask for further dry etching (see Supplementary Materials S4). It is worth mentioning that the inner diameter of the donut pattern defines the diameters of the nanodisks. Then, the sample was developed with $n$-amyl acetate for $4\,\mathrm{min}$ and dried gently with nitrogen gas. Subsequently, the sample was etched using dry reactive ion etcher (RIE) with CHF$_3$ gas. A complete etching of WS$_2$ nanodisks was ensured by performing STEM diffraction, atomic resolution STEM imaging, EELS and energy dispersive spectroscopy (EDS) chemical analysis inside the etched donut-shaped patterns.

\subsection{Experimental STEM imaging and EELS}

STEM experiments were performed at $200\,\mathrm{keV}$ and at room temperature on a JEOL Mono NEO ARM 200F instrument. The microscope is equipped with a Schottky field emission gun, double Wien filter monochromator, probe aberration corrector, image aberration corrector, and Gatan Imaging Filter continuum HR spectrometer. HAADF and SE STEM images were acquired with an around $0.1\,\mathrm{nm}$ electron probe. The STEM image shown in Fig. 2(h) was produced by summing a series of 50 images after the series was registered based on cross correlation to correct for rigid image drifts between frames. The resulting summed image was Fourier filtered to further enhance signal-to-noise ratio and the WS$_2$ signal. EDS experiments utilized a double detector system with a collection solid angle of up to $1.6\,\mathrm{sr}$.

EELS experiments were performed with less than $1\,\mathrm{nm}$ spatial resolution, $20\,\mathrm{meV}-40\,\mathrm{meV}$ energy resolution, a $30\,\mathrm{pA}-50\,\mathrm{pA}$ beam current, a $21\,\mathrm{mrad}$ electron probe convergence angle, and a $10\,\mathrm{mrad}$ EELS collection angle. The EEL spectra shown in Figs. 3(a)-(c) were acquired from just outside the edge of the WS$_2$ nanodisks and are the sum of a series composed of $160,000$ spectra using a $5\,\mathrm{meV}$ dispersion. Before summing the spectral series, a high quality dark spectrum reference was subtracted from each spectra and the spectra were energy-aligned based on the zero energy loss peak position. After summing, the spectra were processed with the Richardson-Lucy deconvolution algorithm using a reference spectrum acquired from the SiN window and only 5 iterations to ensure no introduction of known artifacts \cite{Gloter2003,Colliex2016}. Deconvolution was used to reduce intensity from the zero-loss peak tail in the energy region of interest. Then the remaining tail of the zero-loss peak was removed by fitting each spectrum in the $0.3\,\mathrm{eV}-0.5\,\mathrm{eV}$ range of interest to a power law function, extrapolating that function and subtracting it from the spectrum.

The experimental EEL line profiles shown in Fig. 4 were each produced from three spectrum images of the same sample area. Each spectrum image consisted of $50 \times 3500$ spatial pixels and $2048$ energy channels. To enhance signal-to-noise ratio, each spectrum image was binned by $50$ in the $x$ direction and by $7$ in the $y$ direction to produce a line profile, and the three profiles from each area were summed after the data processing mention below. The spectra were processed with high quality dark spectrum reference subtraction, energy-alignment, Richardson-Lucy deconvolution, and zero-loss peak tail removal using the same details as the spectra mentioned above. Additionally, the EEL line profiles were normalized based on their total integrated EEL signal. The EEL intensity in the as-acquired data is modified by scattering sources other than the low-loss optical and electronic signals that we are interested in. For example, WS$_2$ elastic diffraction scatters a significant amount of electrons outside of the spectrometer collection aperture, significantly reducing all the EEL signals when the electron beam passes through the WS$_2$ nanodisk as compared to when it passes outside the nanodisk. Additionally, variations in the morphology and composition of the residual resist modify the spatial variations in EEL signal intensity. Normalizing each spectrum in the line profile to its own total integrated EEL signal counteracts some of these effects and makes them more comparable to simulations. The spectrum images were acquired in dual EELS mode with a $15\,\mathrm{meV}$ dispersion to enable the collection of EELS data over a larger range of energies and increase the inelastic energy range of the normalization process. Splicing the two dual EEL spectra together allows each spectra to extend to around $50\,\mathrm{eV}$ while still resolving the disk modes and anapole states in the $0.5\,\mathrm{eV}-2.5\,\mathrm{eV}$ range.

\subsection{Numerical simulations}

We performed the numerical simulations shown in Figs. 1, 3 and 4 using the Radio Frequency Module of COMSOL Multiphysics software \cite{COMSOL}. This module solves Maxwell's equations in the frequency domain based on the Finite Element Method (FEM). The nanodisk was modeled as a cylindrical structure of variable radius $R$ and thicknesses $d=55\,\mathrm{nm}$ (Fig. 1) and $d=70\,\mathrm{nm}$ (Figs. 3 and 4). The symmetry axis of the nanodisk is parallel to the $z$-direction and the center of the nanodisk was located at coordinates $(0,0,d/2)$. The probing electron traveling in the $z$-direction was modeled as a line current (along the $z$-direction) located at a distance $b$ (impact parameter) with respect to the nanodisk center. For numerical calculations shown in Figs. 1 and 3, we located the electron beam trajectory outside the nanodisk at an impact parameter $b=1.1\times R$. For simplicity, numerical calculations shown in Fig. 1 were performed without considering any substrate whereas in Figs. 3 and 4 the nanodisk is on top of a $50\,\mathrm{nm}$ thick substrate layer characterized by the permittivity of SiN $\varepsilon_{\mathrm{SiN}}=4.1853$. To ensure numerical convergence of the calculated electron energy loss probability $\Gamma(\omega)$, the complete structure (electron beam and nanodisk) was embedded in a homogeneous box filled with air of depth, width and height equal to $L_{\mathrm{PML}}=12\times R$. We use perfectly matched layers (PML, with thickness equal to $0.1\times L_{\mathrm{PML}}$) for the boundaries of the simulation box, free triangular elements for the the nanodisk mesh and free tetrahedral elements for all other structures. The length of the line current that models the electron beam is equal to $L_{\mathrm{PML}}$. Material permittivities and further details on EEL probability calculations are provided in Supplementary Materials S1.


\section*{Acknowledgements}

C.M.-E. thanks J. Olmos Trigo for fruitful discussions. C.M.-E. and J.A. further acknowledges grant no. IT 1526-22 from the Basque Government for consolidated groups of the Basque University. R.H acknowledges support from the Spanish Ministry of Science and Innovation under the Mar\'ia de Maeztu Units of Excellence Program (CEX2020-001038-M/MCIN/AEI/10.13039/501100011033) and Grant PID2021-123949OB-I00 funded by MCIN/AEI/10.13039/501100011033 and by ``ERDF A way of making Europe''. A.B.Y. acknowledges funding from the Swedish Research Council (VR, under grant No. 2020-04986). T.O.S. acknowledges funding from the Swedish Research Council (VR, Research environment grant No. 2016-06059). A.B.Y., B.M., E.O., and T.O.S. acknowledge funding from the Knut and Alice Wallenberg Foundation (KAW, under grant No. 2019.0149). D.G.B. acknowledges financial support from the Ministry of Science and Higher Education of the Russian Federation (Agreement No. 075-15-2021-606), Russian Science Foundation (grant No. 21-72-00051), and BASIS Foundation (grant No. 22-1-3-2-1). This work was performed in part at the Chalmers Material Analysis Laboratory (CMAL).

\section*{Author contributions}

C.M.-E., D.G.B., and T.O.S. conceived the idea. B.M. fabricated the samples under the supervision of T.O.S. A.B.Y. performed EELS measurements under the supervision of E.O. C.M.-E. performed numerical calculations under the supervision of R.H. and J.A. C.M.-E., D.G.B., J.A., and T.O.S. developed the theoretical model to describe the anapole-exciton system and analyzed the data. C.M.-E., A.B.Y., and D.G.B. wrote the manuscript with contributions from all authors. E.O., J.A., and T.O.S. supervised the project.

\bibliography{scibib}

\end{document}